\documentclass[multphys,vecphys]{svmult}

\usepackage{makeidx}         
\usepackage{graphicx}        
\usepackage{multicol}        
\usepackage[bottom]{footmisc}

\newcommand{\Sauron}{\texttt{SAURON}}

\makeindex             

\begin{document}

\title*{The kinematics of Core and Cusp galaxies: comparing HST imaging and integral-field observations}

\titlerunning{The kinematics of Core and Cusp galaxies}

\author{J. Falc\'on-Barroso\inst{1}
\and
R. Bacon\inst{2}
\and
M. Cappellari\inst{3}
\and
R.~L. Davies\inst{3}
\and
\newline
P.~T. de Zeeuw\inst{4}
\and
E. Emsellem\inst{2}
\and
D. Krajnovi\'{c}\inst{3}
\and
H. Kuntschner\inst{5}
\and
\newline
R.~M. McDermid\inst{4}
\and
R.~F. Peletier\inst{6}
\and
M. Sarzi\inst{7} 
\and
G. van de Ven\inst{8}}

\authorrunning{Falc\'on-Barroso et al.}

\institute{European Space and Technology Centre, Keplerlaan 1, 2200~AG Noordwijk, The Netherlands
\texttt{jfalcon@rssd.esa.int}
\and
Universit\'e de Lyon 1, CRAL, Observatoire de Lyon, 9 av. Charles Andr\'e, 69230 Saint-Genis Laval, France
\and
Sub-Department of Astrophysics, University of Oxford, Denys Wilkinson Building, Keble Road, Oxford OX1~3RH, United Kingdom
\and
Sterrewacht Leiden, Universiteit Leiden, Postbus 9513, 2300~RA, Leiden, The Netherlands
\and
Space Telescope European Coordinating Facility, European Southern  Observatory, Karl-Schwarzschild-Str.~2, 85748 Garching, Germany
\and
Kapteyn Astronomical Institute, University of Groningen, NL-9700 AV Groningen, The Netherlands
\and
Centre for Astrophysics Research, University of Hertfordshire, Hatfield, Herts AL10~9AB, United Kingdom
\and
Institute for Advanced Study, Einstein Drive, Princeton, NJ~08540, USA}

\maketitle

\begin{abstract}
In this proceeding we look at the relationship between the photometric nuclear
properties of early-type galaxies from Hubble Space Telescope imaging and their
overall kinematics  as observed with the \Sauron\ integral-field spectrograph.
We compare the inner slope of  their photometric profiles and the Slow/Fast
rotator classes, defined by the amplitude of  a newly defined $\lambda_R$
parameter, to show that slow rotators tend to be more massive systems and
display shallower inner profiles and fast rotators steper ones. It is important
to remark, however, that there is not a one-to-one relationship between the two
photometric and kinematic groups.
\end{abstract}

\section{Introduction}
\label{sec:1}
\begin{figure}[t]
\begin{center}
\includegraphics[width=0.6\linewidth]{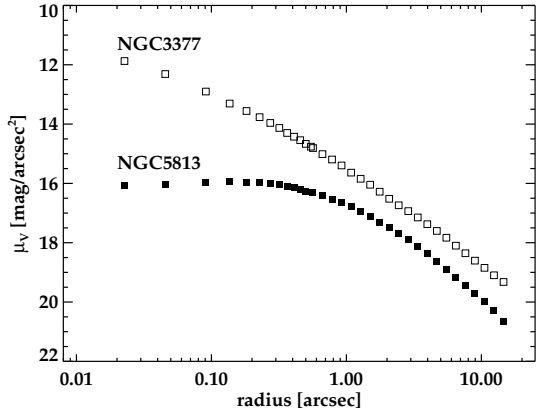}
\end{center}
\caption{An illustration of a core and a cuspy nuclear profiles of two galaxies 
in the \Sauron\ sample from HST V-band observations. Data from \cite{lauer05}.}
\label{fig:1}
\end{figure}

The study of the galactic nuclei in early-type galaxies has played a
fundamental role in our understanding of how galaxies form and evolve. The
arrival of high resolution instrumentation with the Hubble Space Telescope
(HST) in the early 90's opened a new window into the analysis of the nuclear
properties of these systems. One of the major discoveries by  HST in this
respect is the {\em apparent} dichotomy in the nuclear structural properties of
these galaxies (see Fig.\ref{fig:1}). First highlighted by \cite{ferrarese94}
and soon after extended by \cite{lauer95}, the core/cusp properties of
early-type galaxies have been, and still are, under severe scrutiny  by many
groups around the world (e.g. \cite{ravin01,rest01}). Recently the existence of a
dichotomy has been questioned by \cite{ferrarese06} from ACS observations as
part of the ACS Virgo Cluster Survey \cite{cote04} (see also A. Jord\'an
contribution in these proceedings). The topic, however, is far from settled as
illustrated by the numerous recent papers debating the existence of the two
classes (e.g. see \cite{ferrarese07,lauer07} and references therein for an
in-depth discussion on the subject). Independently of the debate, it
is unquestionable that HST is the only current facility in the world that can
allow us to revisit this issue.\looseness-2

Almost in parallel to the progress made by HST, ground-based integral-field
spectroscopy slowly started to emerge allowing us to look at galaxies in a
different new way. With the first observations in 1999, the \Sauron\ project
\cite{dezeeuw02} has been one of the pioneers in the exploitation of this
technology to study the kinematical properties of galaxies. In this
contribution we made use of one of the latest results from our survey
to connect the nuclear photometric properties of early-type galaxies with their 
overall level of rotation. A more detailed analysis of the results presented
here is the matter of study in \cite{emsellem07}.

\section{Core and Cusp galaxies in the \Sauron\ sample}

There are 33 galaxies in the \Sauron\ sample of 48 elliptical and lenticular galaxies
with known values for their inner profile slope, which separates them in "core"
and "power-law" galaxies \cite{faber97,rest01,ravin01,lauer05}. From the point of view
of their stellar kinematics (see \cite{emsellem04}) early-type galaxies appear
in two broad flavours, depending on whether they exhibit clear large-scale
rotation or not. We measure the level of rotation via a new parameter
($\lambda_R$) and use it as a basis for a new kinematic classification that
separates early-type galaxies into slow and fast rotators in \cite{emsellem07}.
We have defined the new quantity $\lambda_{R}$:

\begin{displaymath}
\label{eq:sumLambda}
\lambda_R \equiv \frac{\langle R \, \left| V \right| \rangle }{\langle R \, \sqrt{V^2 + \sigma^2} \rangle}\, ,
\end{displaymath}

that measures the amount of specific (projected) angular momentum from the
velocity maps. The parameter has been defined such that is insensitive to small
features in the maps, and therefore provides a robust measurement of the global
rotation. As we go from galaxies with low to high $\lambda_{R}$ values, the
overall velocity amplitude naturally tends to increase. More importantly,
there seems to be a change in the observed stellar velocity structures.

\begin{figure}[t]
\includegraphics[width=\linewidth]{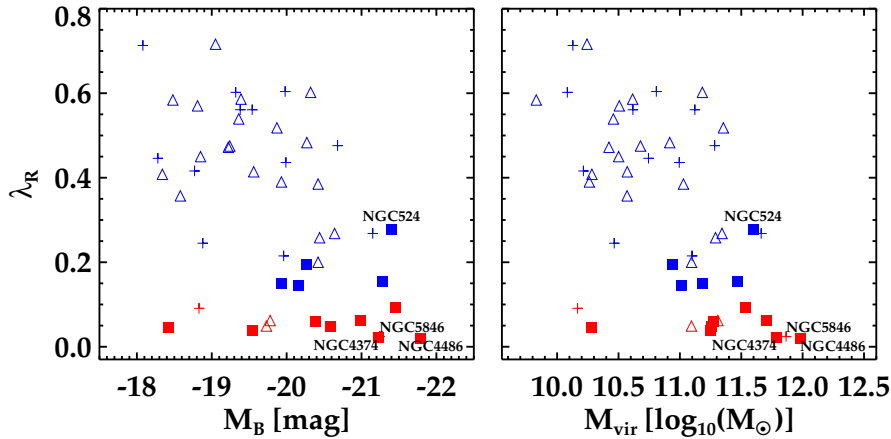}
\caption{$\lambda_{R}$ versus absolute magnitude $M_B$ (left panel) and virial
mass M$_{vir}$ (right panel) for the 48 E and S0 of the \Sauron\ sample. In both
panels, symbols correspond to the inner slope classification
\cite{faber97,rest01,ravin01,lauer05} with power-laws as open triangles, cores as
filled squares, and crosses indicating galaxies for which there is no published
classification. Slow rotators are coloured in red and fast rotators in blue.}
\label{fig:2}
\end{figure}

In the left panel of Fig.~\ref{fig:2}, we show the distribution of
$\lambda_{R}$  as a function of absolute magnitude $M_B$. The three slowest
rotators  (NGC\,4486, NGC\,4374, NGC\,5846) are among the brightest galaxies in
our sample with $M_B < -21$~mag. Other slow rotators tend to be bright but are
spread over a wide range of absolute magnitude. Most fast rotators are fainter
than $M_B >-20.5$~mag. In terms of core/cusp distribution, we find that most
slow rotators are core galaxies while most fast rotators display power-law
profiles. Interestingly there are no core galaxies with $\lambda_{R} > 0.3$,
although the inclined core galaxy NGC\,524 would probably have a very high
$\lambda_{R}$ value if seen edge-on. The general behavior that core galaxies have
lower $\lambda_{R}$ than cusp galaxies is expected, since both classifications
show trends with total luminosity,  with brighter members tending to be core
galaxies, and lower luminosity ones  having power-law profiles \cite{faber97}.
Indeed, all galaxies with  $\lambda_{R} > 0.3$ have $M_B > -20.7$. It is
important to notice though that there is not a one-to-one correspondance
between the different photometric and kinematic groups, since we find both
"power-laws" in slow rotators and  "cores" in fast rotator. One can appreciate
in the figure that there is a domain in luminosity and also in mass where both
cusp and core galaxies coexist.

In the right panel of Fig.~\ref{fig:2}, we show the same trends but now as a
function of the total mass of the galaxies. We have derived the mass values
assuming it approximates to the virial mass derived from the best-fitting 
M$_{vir}- \sigma$  relation presented in \cite{cappellari06}. The figure
displays a trend with $\lambda_{R}$ such that smaller values are found in more
massive galaxies. The three slowest rotators are in the high range of M$_{vir}$
with values above $10^{11.5}$~M$_{\odot}$. There is a clear overlap in mass
between fast and slow rotators for M$_{vir}$ between $10^{11}$ and
$10^{11.5}$~M$_{\odot}$. However, all slow rotators, have M$_{vir} >
10^{11}$~M$_{\odot}$, whereas most fast rotators have M$_{vir} <
10^{11}$~M$_{\odot}$, lower masses being reached as the value of $\lambda_{R}$
increases.

The lack of one-to-one relationship between the photometric and kinematic
classes is probably highlighting the complex merging histories these galaxies
go through in their evolution. Based on the studies showing the incidence of
gas in our sample \cite{sarzi06,emsellem07}, it is likely that dissipation is
one of the dominant factors in the way galaxies in the different classes
evolve. In a forthcoming paper (Falc\'on-Barroso et al., in preparation) we
will combine HST and groung-based imaging with our integral-field observations
to extend the analysis presented in this contribution, and to investigate 
in more detail the links between core/cups and slow/fast rotators in our sample 
of galaxies.



\printindex

\begin{thebibliography}{99.}

\bibitem{cappellari06} Cappellari M., et al., 2006, MNRAS, 366, 1126 
\bibitem{cote04}       C{\^o}t{\'e} P., et al., 2004, ApJS, 153, 223 
\bibitem{dezeeuw02}    de Zeeuw P.~T., et al., 2002, MNRAS, 329, 513 
\bibitem{emsellem04}   Emsellem E., et al., 2004, MNRAS, 352, 721 
\bibitem{emsellem07}   Emsellem E., et al., 2007, MNRAS, 379, 401 
\bibitem{faber97}      Faber S.~M., et al., 1997, AJ, 114, 1771 
\bibitem{ferrarese94}  Ferrarese L., van den Bosch F.~C., Ford H.~C., Jaffe W., O'Connell R.~W., 1994, AJ, 108, 1598 
\bibitem{ferrarese06}  Ferrarese L., et al., 2006, ApJS, 164, 334 
\bibitem{ferrarese07}  Ferrarese L., et al., 2006, "Black Holes: from Stars to Galaxies - Across the Range of Masses", Proceedings IAU Symposium No. 238, eds. V. Karas \& G. Matt.(astro-ph/0609762)
\bibitem{lauer95}      Lauer T.~R., et al., 1995, AJ, 110, 2622 
\bibitem{lauer05}      Lauer T.~R., et al., 2005, AJ, 129, 2138 
\bibitem{lauer07}      Lauer T.~R., et al., 2007, ApJ, 664, 226 
\bibitem{ravin01}      Ravindranath S., Ho L.~C., Peng C.~Y., Filippenko A.~V., Sargent W.~L.~W., 2001, AJ, 122, 653 
\bibitem{rest01}       Rest A., van den Bosch F.~C., Jaffe W., Tran H., Tsvetanov Z., Ford H.~C., Davies J., Schafer J., 2001, AJ, 121, 2431
\bibitem{sarzi06}      Sarzi M., et al., 2006, MNRAS, 366, 1151 
\end{thebibliography}
\end{document}